\documentclass[twocolumn,preprintnumbers,amsmath,amssymb,floatfix,showkeys,pra]{revtex4}
 \usepackage{graphicx}

\begin{document}

\preprint{copyright {\em Journal of Biomedical Optics}, 2008}

\title{Resonance Raman measurements of carotenoids using light emitting diodes}



\author{S. D. Bergeson}
\author{J. B. Peatross}
\affiliation{Department of Physics and Astronomy, Brigham
 Young University, Provo, UT 84602}

\author{N. J. Eyring}
\author{J. F. Fralick}
\author{D. N. Stevenson}
\author{S. B. Ferguson}
 \affiliation{ Pharmanex Research Institute,
  75 West Center, Provo, UT 84601 USA }
 \homepage{http://www.pharmanex.com/}

\begin{abstract}
We report on the development of a compact commercial instrument for
measuring carotenoids in skin tissue. The instrument uses two light emitting diodes (LEDs) for dual-wavelength excitation and four photomultiplier tubes for multichannel detection.  Bandpass filters are used to select the excitation and detection wavelengths.  The $f/1.3$ optical system has high optical throughput and single photon sensitivity, both of which are crucial in LED-based Raman
measurements. We employ a signal processing technique that compensates for detector drift and error. The
sensitivity and reproducibility of the LED Raman instrument compares
favorably to laser-based Raman spectrometers. This compact, portable
instrument is used for non-invasive measurement of carotenoid
molecules in human skin with a repeatability better than 10\%.
\end{abstract}


\maketitle

\section{INTRODUCTION}
\label{sect:intro}  

Carotenoids are important in maintaining proper human health.  They
protect tissues from oxidative stress that might otherwise lead to
premature macular degeneration \cite{rattner06}, cataract formation,
sun burns \cite{wertz05}, premature skin aging, and basal cell and
squamous cell carcinomas \cite{seifried03, dorgan04}. Carotenoids
protect cellular DNA \cite{porrini00} and play a role in the
recovery of burn patients \cite{horton01}. Reliable monitoring of
carotenoid levels in tissue may aid studies to better understand
how they perform these functions.

Resonance Raman spectroscopy has been demonstrated as a rapid,
non-invasive approach to monitoring carotenoid concentrations in
human tissues \cite{gellermann02a, ermakov04b}. The method's success
is primarily due to the low fluorescence quantum yield of these
molecules. While the Raman scattering cross-section increases
dramatically on resonance, fluorescence from carotenoids remains
low.

Several compact laser-based Raman spectrometers have been reported
\cite{caspers01, frank94, hata00, dorgan04, bernstein98,
gellermann02a, ermakov04b}. Typically these use optical fibers to
deliver laser light to the sample and to collect scattered light.
Grating spectrometers and CCD detectors are used to disperse and
detect the scattered light spectrum. This approach offers wide
spectral coverage, aiding in the identification of several
biomolecules in the sample. Fiber coupling of the laser and
fluorescence collection is convenient because the mode diameter and
numerical aperture of the fiber is a good match for both the laser
and the spectrometer. These systems often require environmentally
stable operating conditions. The performance of the lasers and CCD
detectors are sensitive to temperature variations.  In an unstable
environment, frequent recalibrations are required.

The large resonance Raman scattering cross section makes it possible
to use incoherent excitation sources, such as light-emitting diodes
(LEDs).  These sources can offer higher stability in a wider range
of environmental conditions at a lower cost and in a small form
factor. However, these sources have broad spectral bandwidths, do
not couple well into optical fibers, and are poorly matched to the
slit area and numerical aperture of grating spectrometers. Using a
bandwidth-narrowed LED source with a conventional Raman spectrometer
results in recorded spectra that are one to two orders of magnitude
weaker than those measured using a laser source.

This lower signal level can be overcome in part with sensitive
detectors.  For example, an instrument for measuring carotenoids in
the macula was reported \cite{ermakov05a}.  A photomultiplier tube
(PMT) was used to detect the Raman signal transmitted through a
narrowband interference filter.  By tilting the filter it was
possible to change the passband wavelength enough to determine the
baseline fluorescence level at a nearby wavelength and to subtract
this from the Raman signal.  Because the signal-to-background levels
were relatively high (0.20), changes in the background fluorescence
only had a small influence on these carotenoid measurements.

In this paper we describe an instrument that uses spectrally-filtered LEDs for resonance Raman spectroscopy of carotenoids in skin tissue.  For a narrowband excitation source, the signal to background ratio is 20 times smaller in the skin than in the macula, 1\% being a typical value.  For a spectrally filtered LED, this ratio is even smaller, owing to the 1 nm spectral width of the bandwidth narrowing filter.  Such an excitation source emits only 0.5 mW of optical power, resulting in an order of magnitude less light exposure to the skin tissue and therefore much low signal levels compared with a typical laser-based instrument.

We also describe techniques we developed to overcome the difficulties in using the low-intensity excitation sources.  We use PMTs to measure the skin fluorescence for increased sensitivity.  To compensate for both short- and long-term drifts in detector sensitivity inherent to PMTs, we exploit a dual wavelength excitation method. These detectors are not stable enough to use traditional shifted excitation Raman difference spectroscopy (SERDS) \cite{brown07}.  Because the signal-to-background ratio is so small, minor variation in the detector gain on the one minute time scale can compromise measurements of the Raman signal.  Rather than subtract, we divide the spectra from the two excitation sources.  This cancels detector gain variation on a one second time scale.  In this regime it is also essential to control for spatial and optical polarization differences from the two excitation sources.  If not controlled, either of these can introduce errors comparable to or larger than the Raman line.  These techniques are incorporated into a compact commercial instrument with good measurement repeatability under widely varying ambient conditions.  We demonstrate calibration stability over a period of nine months, and it seems likely that much longer stability times can be maintained.

\section{CAROTENOID SPECTROSCOPY}

The optical properties of carotenoids have been studied extensively
over the past few decades. A variety of fluorescence, excited state
absorption, and Raman methods have clarified carotenoid energetics,
and given insight into how these molecules perform their various
chemical roles \cite{sue88, polivka01, yoshizawa03, polivka04}. In
skin tissue, lycopene and $\beta$-carotene are the dominant
carotenoid molecules \cite{ermakov04b}.  They both have a backbone
of nine conjugated carbon bonds, with additional conjugation into
ionone rings on each end.

Optical excitation of carotenoids occurs in the blue-green spectral
region, between 400 and 500 nm.  The three Raman lines in
$\beta$-carotene are at 1008 cm$^{-1}$ (C$-$CH$_3$ rocking mode),
1159 cm$^{-1}$ (C$-$C stretch mode), and 1525 cm$^{-1}$ (C$=$C
stretch mode), with the latter being the strongest. Because of fast
non-radiative relaxation, spontaneous fluorescence is suppressed,
with a fluorescence quantum yield of only $10^{-5}$ \cite{sue88,
gellermann02a}.  The fluorescence background is dominated by
contributions from other tissue components (lipids, proteins,
melanin, DNA, hemoglobin, etc. \cite{anderson81}). The resulting
Raman signal-to-background ratio is typically 1\%
\cite{ermakov01, darvin05}.

\section{OPTICAL DESIGN}
\label{sec:optdesign}

\begin{figure}[b]
\begin{center}
\includegraphics[angle=270,width=3.4in]{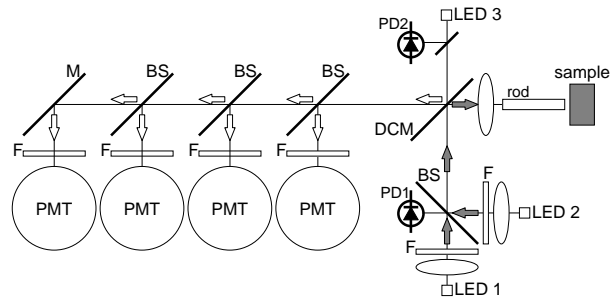}
\end{center}
\caption{\label{fig:optical} Schematic diagram of the optical layout of our
instrument. M = mirror, BS = beam splitter, DCM = dichroic mirror, F
= bandpass filter, L = lens, PD = photodiode.  Bandpass filter
properties are noted in the text.  The optical power at the sample
is less than 0.5 mW in 3.1 mm$^2$.}
\end{figure}

Our application focuses on detection of the 1525 cm$^{-1}$ Raman
line in human skin tissue under resonance excitation conditions. The
optical layout is shown in Fig. \ref{fig:optical}.  A blue LED is
used as the excitation source.  Its nominal 25 nm (FWHM) spectral
width is narrowed to 0.8 nm using a bandpass filter at 473 nm.  This
light reflects from a dichroic mirror and is focused onto the
sample.  Fluorescence and the Raman signal from the sample passes
back through the dichroic mirror.  A series of partially reflecting
beam splitters divides the optical signal into four channels. Light
in each channel passes through an additional 1 nm bandpass filter
and is detected using a photomultiplier tube (PMT). The four 1-nm detection bandpass filters have centers spaced equally on 2 nm intervals.  The center
wavelengths of the detection filters are chosen so that one of the
filters is shifted 1525 cm$^{-1}$ from the excitation source, with
one filter 2 nm to the red and two filters to the blue.  For
excitation at 473.0 nm, the detection filters are at 505.8 nm, 507.8
nm, 509.8 nm, and 511.8 nm. The 509.8 nm filter is centered on the
Raman line.  The other filters characterize the background
fluorescence level.  The PMT detectors have good quantum efficiency at these wavelengths and provide single-photon sensitivity.

Using four detector channels allows us to build a second Raman
spectrometer into the same optical system.  A second excitation
source at 471.3 nm generates a Raman line at the 507.8 nm filter. As
before, we have one detection filter centered on the Raman line
straddled by three ``background'' filters to help establish the
baseline.  With this redundancy, we can implement a measurement
methodology that removes sensitivity to drifts in the PMT efficiency
and gain, as described in Sec. \ref{sec:dsrs}.  Special care must be
taken to minimize differences in the spatial structure between the
two excitation sources, and to control polarization in the
excitation and detection optics, as described in the next section.

\section{DIVIDED SHIFTED RAMAN SPECTROSCOPY}
\label{sec:dsrs}

Using LEDs with interference filters gives us the flexibility
to choose the excitation wavelength more or less arbitrarily.  However,
the corresponding Raman lines and the overall signal levels are
weaker than that obtained under laser excitation.  Using 0.8 nm bandwidth
filters and $f/1.3$ optics results in typically 0.5 mW of optical
power focused to a 2 mm diameter spot.  Because the spectral
widths of the LED excitation
sources are much broader than a laser source, the signal
to background ratio is further reduced.  We compensate for these
disadvantages by using a high throughput optical system with
high sensitivity PMT detectors.

While PMTs have excellent sensitivity, their signal gain can be influenced by temperature variations.  In the analog mode, the gain variation can be as much as 20\% over a few hours as the PMTs warm up.  Because the carotenoid signal is so
small compared to the background, this seemingly slow
drift translates into roughly one percent of the Raman peak signal
per five seconds of measurement time.  For measurement times approaching
100 seconds, this is a 20\% variability in successive measurements.

Because of this, standard flat fielding techniques, such as dividing the measurement from a spectrally smooth sample into subsequent measurements, do not work.  Variations could be minimized by stabilizing the system's environmental conditions. However, stabilizing PMT response at the $10^{-4}$ level is challenging, especially for a compact and portable device intended to operate in variable ambient conditions. Instead we use our two shifted spectral measurements described in Sec. \ref{sec:optdesign} to effectively `flat field' our detectors on a fast time scale while simultaneously extracting the Raman signal. We call our method `divided shifted Raman spectroscopy' (DSRS).

\begin{figure}[b]
\begin{center}
\includegraphics[width=3.4in]{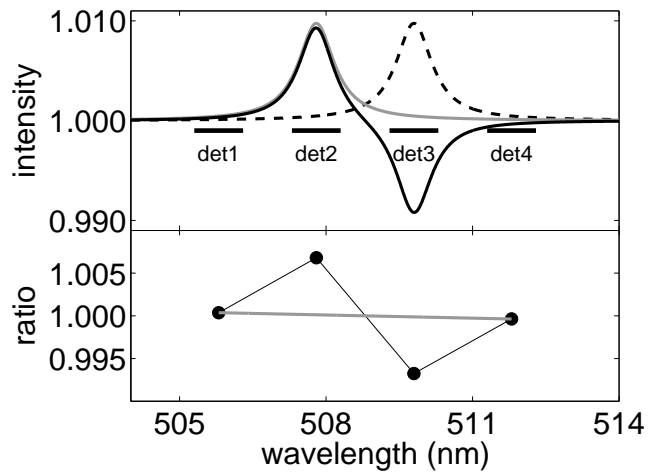}
\caption{\label{fig:ratplot} Top panel: the simulated Raman line
with excitation at 473.0 nm (dashed line) and 471.3 nm (gray line)
and the ratio of the two (black line). The short horizontal black
lines represent the spectral pass bands of the analysis filters.
Bottom panel: the values of the ratio of the two Raman signals in
the top panel as measured by the four detectors, plotted versus
filter wavelength (black circles) and the baseline (gray line).  In
skin measurements, the background has both slope and curvature (not
shown in the figure), although the curvature is always small.}
\end{center}
\end{figure}

For illustrative purposes,
two simulated Raman signals are shown in the top panel of Fig.
\ref{fig:ratplot}. The spectral windows measured by each of the four
detection channels are also shown. The dashed line is the Raman
signal using 473.0 nm excitation. The Raman peak appears in the
third detector and the other three detectors record background
levels. The gray line shows the Raman signal using 471.3 nm
excitation. In this case, the Raman peak appears in the second
detector and the other three detectors record background levels. The
black line shows the ratio of the two spectra. The lower panel of
Fig. \ref{fig:ratplot} shows the ratio of the optical signal
measured by each detector. Note that all effects due to differences
in the detector gain and sensitivity and fluorescence collection
geometry divide out.

\begin{figure}
\begin{center}
\includegraphics[width=3.4in]{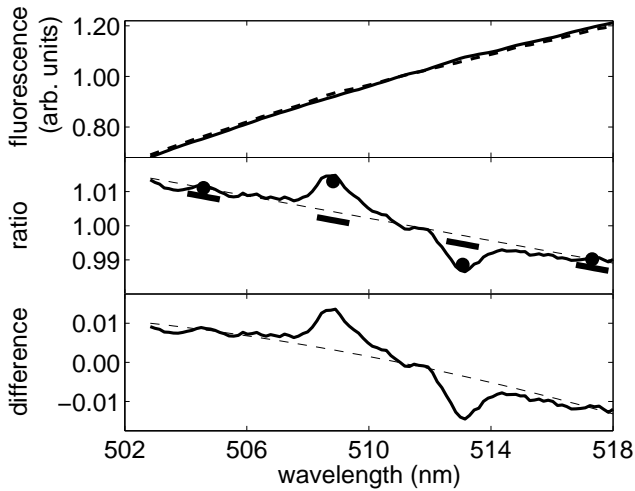}
\end{center}
\caption{\label{fig:dsrs} Top panel: Fluorescence
measurements on a hand showing the 1525 cm$^{-1}$ Raman line using
lasers at 473 nm (dashed line) and 476
nm (solid line).  Middle panel: The ratio of the two spectra from the top panel (solid line) and the baseline fit (dashed line).  The thick solid lines show the 1 nm bandwidth spectral windows used in the data analysis.  The solid circles are plots of the $x_i$'s and $y_i$'s (see Eq. \ref{eqn:array}).
Bottom panel: The difference of the two spectra from the top panel (solid line) and the baseline fit (dashed line) using a similar data analysis approach as in the middle panel.
The excitation intensity was 5 mW
focused to a 0.5 mm spot diameter.  Note that the separation in the laser wavelengths used in this measurement is wider than used in the LED units.}
\end{figure}

To further illustrate the measurement approach of our LED instrument, we performed similar measurements using two laboratory lasers and a conventional grating spectrometer. In this case, there is far more light available for detection than in the LED instrument.  We measured the 1525 cm$^{-1}$ Raman
line excited by a doubled Nd:YAG laser at 473 nm and by an argon-ion
laser at 476 nm.  Typical fluorescence measurements are shown in Fig. \ref{fig:dsrs}. The top panel of Fig. \ref{fig:dsrs} shows spectral measurements of light
collected from a hand with the two excitation sources.  The middle panel
shows the 473/476 fluorescence ratio.  The short black lines show equally spaced wavelength intervals, similar to what might be transmitted by bandpass filters.  The black circles show the average signal over the analysis wavelength ranges.  The thin dashed line shows a parabolic fit to the background.  As is typically the case, the quadratic term is small.

There are similarities in this approach and in
the more conventional and successful
`shifted excitation Raman difference spectroscopy' (SERDS)
approach \cite{brown07}.  We can imagine that each of the two spectra shown in the top panel of Fig \ref{fig:dsrs} is represented by a smooth function of wavelength plus some small variation.  The fluorescence signal with 473 nm excitation can be written as ${\cal S}_{473}(\lambda) = f(\lambda) + g_1(\lambda)$.  Similarly, the fluorescence signal with 476 nm excitation can be written as ${\cal S}_{476}(\lambda) = f(\lambda) + g_2(\lambda)$.  The ratio of these two spectra is
\begin{equation} \label{eqn:rat}
\frac{{\cal S}_{473}}{{\cal S}_{476}} =
\frac{f(1+g_1/f)}{f(1+g_2/f)} \approx
1 + \frac{g_1}{f} - \frac{g_2}{f}.
\end{equation}
If the function $f$ is relatively flat over the two Raman lines represented by the functions $g_1$ and $g_2$, division gives the difference of the scaled Raman signals centered about 1.  The fact that $f$ is not flat over the two lines and that $g_1$ and $g_2$ contain more spectral data than just the Raman lines results in a background that is not perfectly flat.  Except for an offset, the ratio (middle panel of Fig. \ref{fig:dsrs}) and the difference (bottom panel of Fig. \ref{fig:dsrs}) are nearly identical.
The important distinction for LED/PMT measurements is that our detectors require rapid flat-fielding for measurements accurate at a level of a few times $10^{-4}$ of the fluorescence level, which translates into a few percent accuracy in the measurement of the Raman signal strength. In contrast, SERDS would suffer from drift errors on the order of $10^{-2}$ of the fluorescence level in our application, translating into 100\% variation in the Raman signal strength measurements.

We establish the baseline in the DSRS spectrum using the endpoints,
which are always ``background'' measurements.  We constrain
the baseline by assuming that the ratio measurement in detector 2 is
as far above the baseline as the ratio measurement in detector 3 is
below it.  These assumptions yield a coupled system of equations for
the background and signal strength:
\begin{eqnarray}
y_1 & = & a + bx_1 + cx_1^2 \nonumber \\
y_2 & = & a + bx_2 + cx_2^2 + d \label{eqn:array} \\
y_3 & = & a + bx_3 + cx_3^2 - d \nonumber \\
y_4 & = & a + bx_4 + cx_4^2 \nonumber
\end{eqnarray}
\noindent where $y_i$ are the ratios of PMT measurements of the two
excitation sources, and the $x_i$ are the interference filter
passband centers.  The parameters $a$, $b$, and $c$ are the
coefficients of the parabola that best represents the baseline. The
parameter $d$ is magnitude of the difference between the Raman
signal and the baseline on this relative scale. Note that these
equations are derived assuming that $d$ is small (e.g., $d\lesssim
0.01$ as for skin measurements).

When $d$ approaches
unity the DSRS method is inapplicable.
This is the
regime in which standard Raman measurement techniques can be
implemented.  However, even high concentration samples with
$d \sim 0.3$ can be reliably measured using our
technique because we can include higher-order corrections
to determine $d$ in Eqs. \ref{eqn:rat} and \ref{eqn:array}.

The parameter $d$ is put on an absolute scale by multiplying it by
the PMT voltages used in the denominator of $y_i$.  To achieve
good reliability, the detector stability needs only to be maintained
(or measured) at the few percent level, which is much simpler than
the $10^{-4}$ required without using the DSRS technique.  We
characterize the PMT response to 1\% percent using a calibration LED
with known brightness preceding each measurement. Absolute
calibration is achieved by measuring the Raman response of a sample
with known carotenoid concentration.

With a different set of assumptions, this scheme could be
implemented with fewer detectors and we have tested these
approaches. We find that instruments with two or three detectors do
not give the repeatability or reliability of the four detector
design, largely because the fluorescence background has both slope
and curvature.  The parabolic analysis in Eq. \ref{eqn:array} makes
it possible to derive a curved baseline with high reliability. A
configuration with a greater number of detectors could be used, but
the additional redundancy does not seem to offer noticeably better
machine performance than what is achieved using four.

In principle, the division analysis method could be used with
conventional CCD detection devices.  It would be most appropriate
when the Raman spectrum was dominated by background fluorescence.
The experimental realization of the division technique in this paper
requires a simple Raman spectrum, such as in resonance measurements.
The background measurement points need to be free of any Raman
lines.  Our implementation of the DSRS technique depends on prior knowledge of
detailed spectral features.

When using two independent excitation sources, it is important to
make sure that the spatial profile of the two sources is identical.
Variations in this profile result from both the structure of the
LEDs themselves, but also in the interference filters used to narrow
their output spectrum.  We do this using a 1 mm diameter, 20 mm long glass rod.  The material was chosen to have a high index of refraction and a low fluorescence level.  This rod acts somewhat like a very large diameter optical fiber.  The  multiple internal reflections remove residual differences in the spatial profile of the two excitation sources.
It is also important to make sure
the optical polarization of the two sources is identical, and to
make sure that the polarization is identical at the beamsplitter preceeding each PMT.  The
Raman signal has the same polarization as the excitation source.
This is not true for the background fluorescence.  If the collected fluorescence at each PMT is not identical, a markedly different
signal-to-background ratio will be measured and the derived Raman signal strength will be compromised.  Special care must be taken because the Fresnel reflection coefficients at 45 degrees angle of incidence are typically quite different for the two polarizations in commercial beamsplitters.

\section{BENCHMARK COMPARISON TO LASER-BASED UNITS}

\begin{figure}[b]
\begin{center}
\includegraphics[width=3.4in]{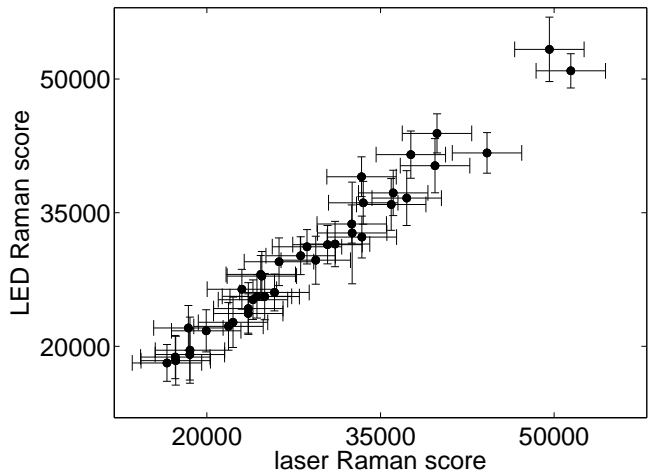}
\caption{\label{fig:laser} Comparison of laser-Raman and LED-Raman
carotenoid measurements on 38 people.  The absolute
calibration of the LED instrument was chosen to match the laser-Raman instrument.  The vertical error bars are
the 1$\sigma$ standard deviations for 20 measurements on two LED
Raman instruments over five days.  The horizontal error bars are
typical 1$\sigma$ standard deviation measurements for the laser
Raman instruments. The measurement time is two minutes for the LED
Raman instrument and three minutes for the laser Raman instruments.
To within the measurement uncertainty, this graph shows perfect
correlation between the two machines.  The reduced $R^2$ value of a
straight line fit is 0.96.}
\end{center}
\end{figure}

A few years ago, Pharmanex$^{\mbox{\tiny \textregistered}}$
developed a laser-based Raman instrument for detecting carotenoid
molecules in human skin tissue.  It was based on a larger laboratory
research device \cite{bernstein98, ermakov01, ermakov01a}.  A
fiber-coupled laser was focused onto the hand, and
backwards-directed fluorescence was fiber-coupled into a grating
spectrometer.  The light intensity in the Raman peak was measured in
the spectrometer trace.  Raman measurements
correlate with carotenoid serum levels \cite{ermakov05b} and both
resonance and non-resonance Raman measurement can be used to
identify, quantify, and monitor molecular content in skin tissue
\cite{hammond04, gellermann04,gniadecka04, nijssen02, caspers01,
hata00, xiao05}.

Pharmanex$^{\mbox{\tiny \textregistered}}$ developed an arbitrary scale for quantifying carotenoid levels in skin tissue that is proportional to the Raman signal.  On this scale, the average American has a reading in the neighborhood of 25,000 for
measurements made in the palm of the hand.  The palm has the advantage of being relatively homogeneous and melanin-free.  Typical scores range
from 10,000 to 50,000, similar to the spread of the population seen
in Fig. \ref{fig:laser} and the measurement uncertainty is less than
3,000.

We have performed hundreds of tests comparing the performance of our
units and laser-based Raman spectrometers.  The results of one test
are shown in Fig. \ref{fig:laser}.  For this study, 38 persons
repeatedly measured their antioxidant scores on a laser Raman
instrument and also on LED Raman instruments over a one week time
period. To within the error of the measurements, the two different
kinds of instruments read identically to one another. A
straight-line fit to the data in Fig. \ref{fig:laser} has a reduced
$R^2$ value of 0.96. The LED Raman instruments also show high
repeatability. The vertical errorbars in Fig. \ref{fig:laser} are
the 1$\sigma$ standard deviations of 20 measurements on two LED
Raman units.  The average relative standard deviation (RSD) is 9\%.
This is the 1$\sigma$ variation in a person's score divided by their
mean score and averaged over the 38 persons in this study. The
typical RSD is somewhat lower for scores in the 25000 to 50000
range.

The LED Raman instruments are accurate in spite of their low
intensity illumination of the skin.  The optical power on the skin
is less than 0.5 mW in a 3.1 mm$^2$ area.  The resulting maximum
intensity is 0.016 W/cm$^2$, which is 12.5$\times$ lower than the
maximum permissible exposure for laser radiation of comparable
wavelength \cite{ansi}.

\section{LONG-TERM CALIBRATION RELIABILITY}

The LED Raman instruments are designed to have long-term calibration
reliability. The optical system includes a photodiode to measure the
brightness of the emission sources (PD1 in Fig. \ref{fig:optical})
and also an additional LED and photodiode to measure the PMT
sensitivity (LED3 and PD2 in Fig. \ref{fig:optical}).  These items
constitute a set of internal calibration standards for the
instrument. They allow us to correct readings for drifts in
excitation intensity and PMT sensitivity over time that may arise
from temperature variation or component aging.

In normal use, a unit is calibrated at the beginning of a
measurement session using a portable Raman standard.  All of the
PMT, LED, and photodiode values are recorded.  Using the internal
calibration standards, we correct all readings back to the LED
brightness and PMT sensitivities as measured at the time of
calibration.

\begin{figure}[t]
\begin{center}
\includegraphics[width=3.4in]{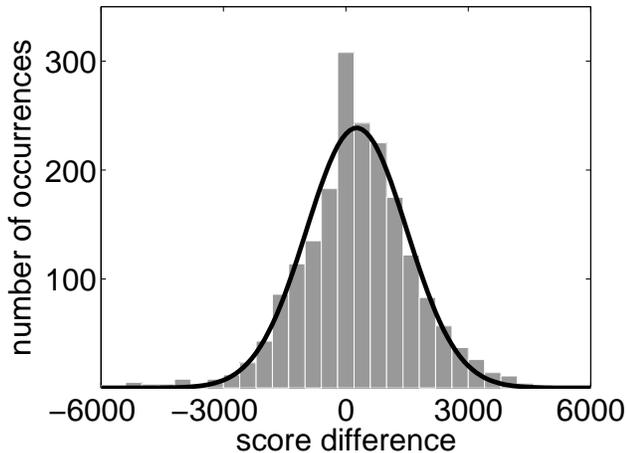}
\end{center}
\caption{\label{fig:hist} Histogram of score differences over a 265
day period.  The x-axis is the difference between the scores
determined using on-site calibration (147 different calibration
events) and the scores determined using the internal calibration
standards.  The 1$\sigma$ width (68.3\% confidence level) determined
by the Gaussian fit to the data (black line) is $\pm1235$.}
\end{figure}

The internal calibration standards are highly robust.  Instead of
using them to refer to conditions at the beginning of a particular
measurement session, they can be used to refer to conditions at a
factory calibration.  In Fig. \ref{fig:hist} we compare the scores
measured over a 265 day period determined in two ways. The first is
the ``normal'' way, using calibrations performed at the point of
measurement.  The second way uses the internal calibration standards
to refer back to a single factory calibration.  The data in Fig.
\ref{fig:hist} is a histogram of differences in the scores
calculated in these two ways. The overall shift in score is
approximately 1\%. The 1$\sigma$ confidence level (68.3\%) is
$\pm1235$. A closer evaluation of the data suggests that most of
these score differences arise from errors in the calibration at the
point of measurement. These data show that the internal calibration
standards are stable over time periods approaching a year, and
stability over much longer time periods seems likely.

\section{CONCLUSION}

We report on the development of a compact and portable Raman instrument
for measuring carotenoid molecules in human skin.  Thousands of
these have been built and are in routine operation.  These units use
high sensitivity detectors and a high throughput optical system,
making it possible to use incoherent light sources for resonance
Raman spectroscopy in a commercial setting. We have shown
repeatability better than 10\% on LED Raman instruments and
excellent correlation with laser-based Raman spectroscopy
measurements.  This work represents an important step forward for
non-laboratory {\em in vivo} Raman spectroscopy using low intensity
excitation.  With the expanding availability of ultraviolet LEDs,
and considering that many important biomolecules have resonances in
the ultraviolet wavelength range, this approach may be useful for
other systems.

\end{document}